\documentclass[12pt,thmsa]{article}
\usepackage{amsfonts}
\usepackage{graphicx}
\usepackage{epsfig}
\usepackage{graphics}

\makeatletter
\newcommand{\row}[1]{\mathord{\buildrel{\lower3pt\hbox{$\scriptscriptstyle\rightarrow$}}\over #1}}

\newcommand{\dyadic}[1]{\mathord{\dyadic@rrow{#1}}}
\newcommand{\dyadic@rrow}[1]{
\begin{picture}(12,12)(-1,0)
\put(-3,12){\makebox(0,0)[t]{$\scriptscriptstyle\downarrow$}}
\put(-3,13){\makebox(0,0)[l]{$\scriptscriptstyle\longrightarrow$}}
\put(5,0){\makebox(0,0)[b]{$#1$}}
\end{picture}
}
\newcommand{\bra}[1]{\bigl\langle #1 \bigr|}
\newcommand{\ket}[1]{\bigl| #1 \bigr\rangle}
\newcommand{\expect}[1]{\left\langle #1 \right\rangle}

\input{tcilatex}

\begin{document}

\date{\today}

\begin{center}
{ Skew Information for a single Cooper Pair Box interacts with  a single cavity Mode}\\[0pt]
\vspace{0.5cm}N. Metwally$^{1,2}$ A. Al-Mannai $^{2}$ and M. Abdel-Aty $^{1,3}$ \\
 $^1$Math. Dept., College of Science, University of
Bahrain, Kingdom of Bahrain\\
$^2$ Math. Dept., Faculty of science, Aswan,  University of Aswan,
Egypt.\\
 $^3$Math. Dept., Faculty of science, Sohag,  University of Sohag, Egypt.\\
 Nasser@Cairo-svu.edu.eg
\end{center}

\begin{abstract}
The dynamics of the skew information (SI)  is investigated for a
single Cooper Pair Box,CPB interacts with a single cavity mode.
The effect of the cavity and CPB's parameters on the SI is
discussed. We show that, it is possible to increase the skew
information to reach its maximum value either by increasing the
number of photons inside the cavity or considering non-resonant
case with larger detuning parameter. The effect of the relative
ratio of Josephson junction capacity and the gate capacity is
investigated, where the number of oscillations of the skew
information increases by decreasing this ratio and consequently
the travelling time between the maximum and minimum values
decreases.
\end{abstract}

pacs{74.70.-b, 03.65.Ta, 03.65.Yz, 03.67.-a, 42.50.-p} 

\section{Introduction}

Cooper pairs  represent one of the most important promising
candidate in the context of quantum information filed. These pairs
are classified as  a physical realization of the solid states
qubit \cite{Yu}. In  quantum teleportation, the generated
entangled  state between the Cooper pair Box CPB and the cavity
mode is used as a quantum channel to perform the original
\cite{benn} quantum teleportation protocol \cite{metwally1}. In
the presence of noise these channels are employed to implement the
original quantum teleportation protocol \cite{haleem}.  The
quantum computational speed of a single Cooper Pair box is
investigated by evaluation the speed of orthogonality\cite{Doaa}.
Moreover, the generated channel between a single CPB and   a
cavity mode is used to send coded information in  the presence of
perfect and imperfect operation during the coding process
\cite{metwally2}. The dynamics of the purity, entropy and the
coherent vector of a single  the Cooper pair interacts with a
single cavity mode is investigated in \cite{metwally3}.

 Recently, skew information
has been used as a test for quantum entanglement , where the Bell
inequality is obtained by means of the skew information. This new
inequality provides an exact test to distinguish entangled from
non-entangled pure states of two qubits \cite{Chen}. Pezz\`{e}
et.al  has introduced a measure of entanglement between
multi-particles in terms of Fisher information \cite{Pezze}. Luo
\cite{Luo} has find a mathematical between the Fisher information,
which many applications in quantum information and  the skew
information. This motivates us to investigate the skew information
for   a single Cooper Pair Box (CPB) interacts with a single
cavity mode..

In this contribution we investigate  the dynamics of the skew
information for a single Cooper pair box  interacts with a single
cavity mode. The effect of the cavity and the Cooper pair box's
parameters on the dynamics of the skew information is discussed.
We show that the skew information increases faster for larger
values of photons inside the cavity or increasing the detuning
parameter between the cavity and the CPB. Also, the effect of the
relative ratio of Josephson junction capacity and the gate
capacity on the behavior of the skew information is investigated

The paper is organized as follows: in Sec. 2, we introduce a brief
discussion on the qubit cavity interaction and its dynamics. Sec 3
is devoted to the skew information, where we review its definition
in the context of quantum information and its relation to the
degree of entanglement. Also, we discuss the dynamics of the skew
information for different values of the CPB and cavaity's
parameters. Finally, the results are concluded in Sec.4.

\section{The Suggested Model}

A single Cooper Pair Box CPB is an example of a qubit with states
$\ket{0}$ and $\ket{1}$ \cite{Bose}. It consists of a small
superconducting island connected to the outside by Josephson
tunnel junction $E_j$ and a gate capacitor $C_J$. A gate voltage
$V_g$ is coupled to the superconducting island through gate
capacitance $C_g$ \cite{Nak,Tsai}.
 This system can be described by  two levels  system with hamiltonian
\begin{equation}
H_{s}=4E_{c}(n-n_{g})^{2}-E_{j}\cos\phi ,  \label{sys1}
\end{equation}%
where, $E_{c}=\frac{1}{2}e^{2}\left( C_{J}+C_{g}\right)$ is the
charging energy, $E_{J}=\frac{1}{2}\frac{\hbar}{e} I_c$  is the
Joesphson coupling energy, $e$ is the charge of the electron,
$n_{g}=\frac{1}{2}\frac{V_{g}}{e}C_g$ is the dimensionless gate
charge, $n$ is the number operators of excess Cooper Pair on the
island and $\phi$ is the phase operator \cite{Mig,Zhang}. The
Hamiltonian of the system (\ref {sys1})  reduces to,
\begin{equation} H_{s}=-\frac{1}{2}B_{z}\sigma
_{z}-\frac{1}{2}B_{x}\sigma _{x},
\end{equation}%
where it is assumed that the  temperature is low enough and
Josephon coupling energy, $E_{j}$
 is much smaller than the charging energy i.e. $E_{j}<<E_{c}$ and
$B_{z}=-\left( 2n-1\right) E_{cl}$, $E_{cl}$ is the electric
energy, $B_{x}=E_{j}$  and $\sigma _{x},\sigma _{y},\sigma _{z}$
are Pauli matrices \cite{Haleem}. If the single Cooper pair Box,
CPB is placed inside a single-mode microwave cavity, then the
Hamiltonian of the system can be written as \cite{Doaa},

\begin{equation}
\mathcal{H}=\varpi a^{\dagger }a+\varpi _{c}\sigma _{z}-g\{\mu -\cos\theta \sigma _{z}+%
\sqrt{1-\nu ^{2}}\sigma _{x}(a^{\dagger }+a)\},  \label{Sys2}
\end{equation}%
where,  $\omega$  is the cavity resonance frequency,
$\omega_c=\sqrt{E_{j}^{2}+16E_{c}^{2}\left( 2n_{g}-1\right) ^{2}}$
 is the transition frequency of the cooper pair qubit, $g
=\frac{\sqrt{C_{j}}}{C_{g+C_{J}}}\sqrt{\frac{1}{2}\frac{\varpi
}{\hbar }e^{2}}$ is coupling strength of resonator to the Cooper
Pair Box,  $\mu =1-n_{g}$, and $\theta =-\arctan
\Bigl(\frac{1}{E_{c}}\frac{E_{j}}{2n_{g}-1}\Bigr)$ is mixing
angle.

 Let us assume that a single Cooper pair, prepared initially an a
superposition state as
$\ket{\psi_c(0)}=\frac{1}{\sqrt{2}}(\ket{e}+\ket{g})$ interacts
with a single cavity mode, prepared initially in a number state,
$\ket{\psi_f}=\ket{n}$. In this case, the initial state of the
total system is defined by
$\ket{\psi_s(0)}=\frac{1}{\sqrt{2}}(\ket{e,n}+\ket{g,n})$. The
time evolution of the initial state vector is given by,
\begin{equation}
\left\vert \psi _{s}(t)\right\rangle =\mathcal{U}(t)\left\vert
\psi _{s}(o)\right\rangle ,  \label{int}
\end{equation}
where, $\mathcal{U}(t)$ is a unitary operator defined by,
\begin{equation}
\mathcal{U}(t)=\mathcal{B}_1\ket{e}\bra{e}+\mathcal{B}_2\ket{e}\bra{g}+\mathcal{B}_3\ket{g}\bra{e}+\mathcal{B}_4\ket{g}\bra{g},
\end{equation}
where,

\begin{eqnarray}
\mathcal{B}_1&=&C_{n+1}-i\delta S_{n+1}, ~\mathcal{B}_2=-iS_n a,
\nonumber\\
\mathcal{B}_3&=&iS_n a^{\dagger}, ~\mathcal{B}_4=c_{n+1}+i\delta
S_{n+1},
\nonumber\\
 C_{n}&=&\cos \Omega\tau\sqrt{(\Delta^{2}+n},\quad S_{n}=\frac{2\lambda }{\sqrt{\Delta
^{2}+4g^{2}n}}\sin\Omega\tau\sqrt{\Delta^{2}+n},\
\nonumber\\
 \Omega&=&\frac{\sqrt{c_j}}{C_j+C_g},\quad
T=\sqrt{\frac{\omega}{2\hbar}} ~  \mbox{ and}~
\Delta=\frac{\delta}{2g}.
\end{eqnarray}
Using Eq.(5), then the state vector (4) becomes,
\begin{equation}\label{CPB}
\ket{\psi_s(t)}=\mathcal{A}_1\ket{e,n}+\mathcal{}_2\ket{g,n+1}+\mathcal{A}_3\ket{e,n-1}+\mathcal{A}_4\ket{g,n},
\end{equation}
where
\begin{eqnarray}
\mathcal{A}_1=\frac{\mathcal{B}_1}{\sqrt{2}},
~\mathcal{A}_2=\frac{\mathcal{B}_1}{\sqrt{2}}\sqrt{n+1},
~\mathcal{A}_3=\frac{\mathcal{B}_1}{\sqrt{2}}\sqrt{n},~
\mathcal{A}_4=\frac{\mathcal{B}_4}{\sqrt{2}}.
\end{eqnarray}
Since, the final state of the initial state vector is obtained,
then we can investigate all the classical and quantum phenomena
associated with this quantum state vector. In this context, we are
interested to investigate the dynamics of the Skew information.

\section{Skew Information}
Skew  information ($\mathcal{S}_I$)  represents  a measure  in
information content of the density operator with respect to a self
adjoint operator \cite{Winger}. Mathematically for a density
operator $\varrho$ and a self- adjoint operator $\mathcal{H}$, the
skew information is defined by
\begin{equation}\label{SW}
\mathcal{S}_I=\frac{1}{2}tr\{\sqrt{\varrho}\mathcal{H}-\mathcal{H}\sqrt{\varrho}\}^2.
\end{equation}
Therefore one can say that $\mathcal{S}_I$ measures the
non-commutativity between $\varrho$ and $\mathcal{H}$ \cite{Luo}.
Moreover, it has been shown that  the skew information can by used
to detect entanglement, where  the Bell inequality is proposed in
terms of $\mathcal{S}_I$ and , it is proved that the inequality
provides an exact test to distinguish entangled from separable
pure states of two qubits \cite{Chen}. In an equivalent form of
 (\ref{SW}), the skew information can be written as,
\begin{equation}
\mathcal{S}_I=tr\{\varrho\mathcal{H}^2\}-tr\{\sqrt{\varrho}\mathcal{H}\sqrt{\varrho}\mathcal{H}\}.
\end{equation}

To investigate the skew information for the Cooper Pair Box
interacts with a cavity mode initially prepared in a number state,
we consider  a set of spin $\frac{1}{2}$ operators for the CPB and
the cavity  as $\sigma_i$ and $\tau_i$ , $i=x,y,z$ respectively.
Now, the skew information for the CPB and the cavity is defined as
\begin{equation}\label{SWC}
\mathcal{S}_I=\sum_{i}\{\Delta\sigma_i^2+\Delta\tau_i^2\},
\end{equation}
where $i=x,y,z$ and $\Delta{A}^2=\expect{A^2}-\expect{A}^2$. Using
(\ref{CPB}) and  (\ref{SWC})  one obtains the  skew information
for the suggested system. However,  for a two -qubit pure system
the skew information and the concurrence  are connected by
\cite{Sun}
\begin{equation}
\mathcal{S}_I=1+\mathcal{C}^2,
\end{equation}
where $\mathcal{C}$ is  the concurrence which quantify the degree
of entanglement between the CPB and the field, where for two
qubits,  the concurrence is calculated in terms of the eigenvalues
$\eta_1,\eta_2,\eta_3,\eta_4$ of the matrix $R=\rho\sigma_y\otimes
\sigma_y\rho^*\sigma_y\otimes\sigma_y.$ It is given by
$\mathcal{C}=max\{0, \eta_1-\eta_2-\eta_3-\eta_4\}$, where
$\eta_1\geq\eta_2\geq\eta_3\geq\eta_4$. For maximally entangled
states concurrence is $1$ while for separable states it is zero
\cite{Woot}.

\begin{figure}[t!]
\begin{center}
\includegraphics[width=25pc,height=14pc]{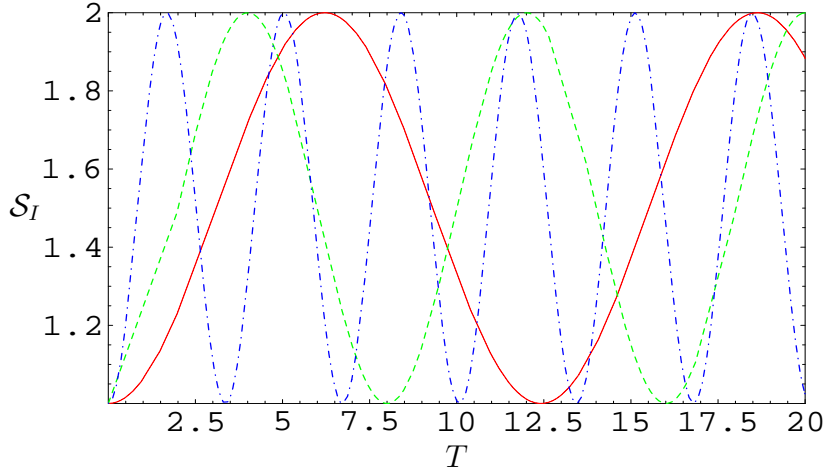} %
 \put(-145,-10){$T$}
  \put(-310,85){$\mathcal{S}_{I}$}
 \end{center}
\caption{The skew information $\mathcal{S}_I$ between the Cooper
pair and the cavity mode for a system initially prepared in
$\ket{n,e}$. The solid , dot and dash dot curves are evaluated for
$\Delta=0.0,0.3$ and $0.9$ respectively. The ratio
$\gamma=\frac{C_j}{C_g}=\frac{1}{4}$ and the number of photon
inside the cavity $n=1$. }
\end{figure}
\begin{figure}
\begin{center}
\includegraphics[width=25pc,height=14pc]{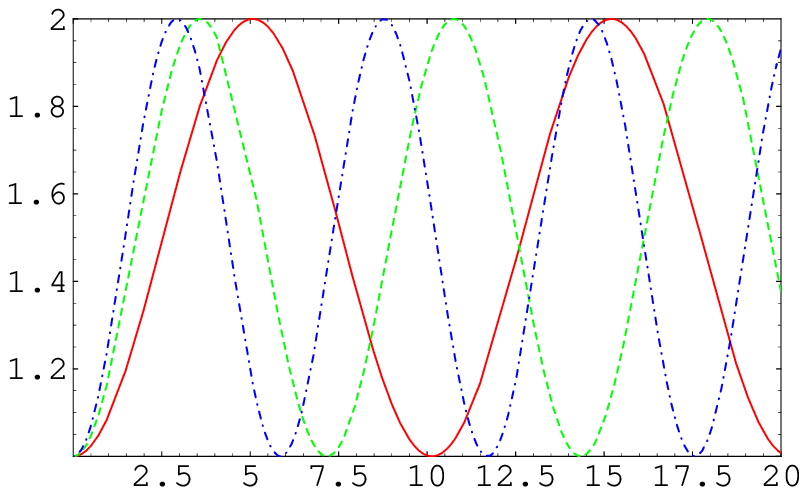} %
 \put(-145,-10){$T$}
 \put(-310,85){$\mathcal{S}_{I}$}
 \end{center}
\caption{The same as Fig.1, but for different values of the photon
inside the cavity. The solid , dot and dash dot curves are
evaluated for $n=2,5$ and $8$ respectively. The ratio
$\gamma=\frac{C_j}{C_g}=\frac{1}{4}$ and the  detuning parameter
$\Delta=0.0$. }
\end{figure}
The dynamics of the skew information for different values of the
detuning parameter is  displayed in Fig.(1), where it is assumed
that the other parameters are constant. It is clear that, for the
resonant case i.e. $\Delta=0.$, the skew information
$\mathcal{S}_I$ increases as $T$ increases to reach its maximum
value for the first time at $T\simeq=7$. However for larger
$\tau$, the skew information decreases smoothly to reach its
minimum value for the first time at $\tau\simeq 12.5$. This
behavior is repeated as the scaled time increases.  For non
resonant case, as example for  ($\Delta=0.3)$, the behavior of the
skew information is similar to that predicted in the non-resonant
case. However, $\mathcal{S}_I$ increases faster and reaches its
maximum value for the first time at $T\simeq 3$. As the
interaction increases the skew information reaches its minimum
values faster than that depicted for resonant case.

\begin{figure}
\begin{center}
\includegraphics[width=25pc,height=14pc]{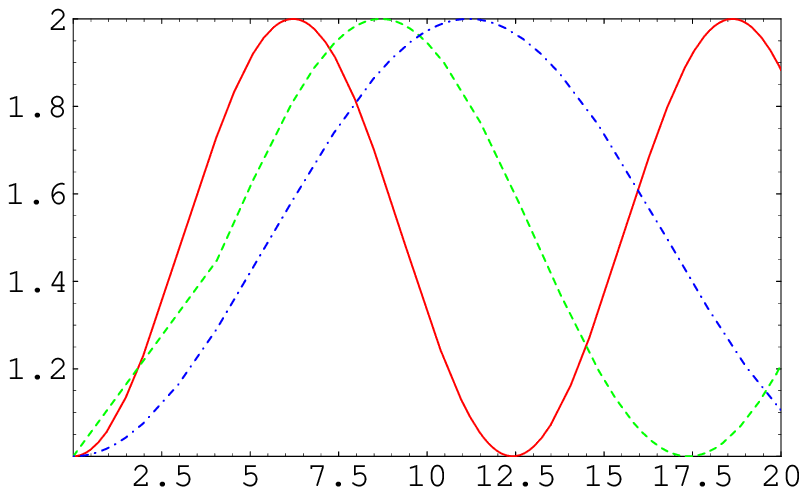}
\put(-145,-10){$T$}
 \put(-310,85){$\mathcal{S}_{I}$}
 \end{center}
\caption{The same as Fig.1, but for different values of the the
ratio $\gamma$. The solid, dot and dash dot curves are evaluated
for $\gamma=\frac{1}{4},\frac{1}{6}$ and $\frac{1}{8}$
respectively. The  number of photons inside the cavity $n=2$, and
the detuning parameter (a) for resonant case i.e .$\Delta=0.0$ (b)
for non-resonant case with $\Delta=0.3$. }
\end{figure}

\begin{figure}
\begin{center}
\includegraphics[width=25pc,height=14pc]{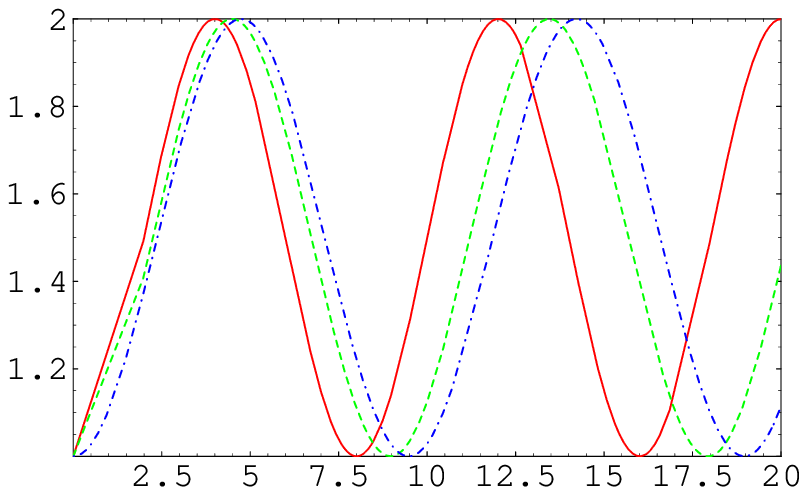}~
 \put(-145,-10){$T$}
 \put(-310,85){$\mathcal{S}_{I}$}
 \end{center}
\caption{The same as Fig.1, but for different values of the the
ratio $\gamma$. The solid, dot and dash dot curves are evaluated
for $\gamma=\frac{1}{4},\frac{1}{6}$ and $\frac{1}{8}$
respectively. The  number of photons inside the cavity $n=2$, and
the detuning parameter (a) for resonant case i.e .$\Delta=0.0$ (b)
for non-resonant case with $\Delta=0.3$. }
\end{figure}

\begin{figure}
\begin{center}
\includegraphics[width=25pc,height=14pc]{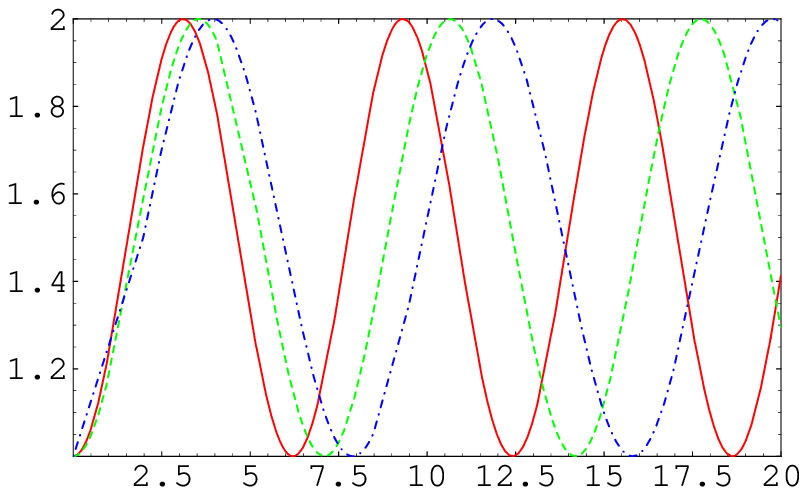} %
 \put(-145,-10){$T$}
 \put(-310,85){$\mathcal{S}_{I}$}
 \end{center}
\caption{The same as Fig.1, but for different values of the ratio
$\gamma$. The solid, dot and dash dot curves are evaluated for
$\gamma=4,6$ and $8$ respectively while the and the number of
photons inside the cavity $n=2$, and the detuning parameter
$\Delta=0.0$. }
\end{figure}

Fig.(2), displays the effect of different numbers of photons
inside the cavity. In this figure, we consider a non-resonant
case, i.e $\Delta=0.0$ and we fix the ratio
$\gamma=\frac{C_j}{C_g}=\frac{1}{4}$. The dynamics of the
$\mathcal{S}_I$ is  similar  to that shown in Fig.(1). However the
number of oscillations depends on the number of photons inside the
cavity. It is clear that, for larger values of $n$, the skew
information reaches its maximum value faster than that evaluated
for smaller number of photons inside the cavity and consequently
$\mathcal{S}_I$ vanishes earlier for larger values of $n$.
Comparing Fig.(1)(solid curve) and its corresponding one in
Fig.(2), we can see that for $n=1$, the skew information reaches
its minimum value at $T=12.5$, while for $n=2$ (solid curve in
Fig.2), the skew information reaches its minimum value at $T=10$.

The effect of the ratio $\gamma$ is displayed in Fig.(3), for
resonant and non-resonant cases, where we fix the other
parameters. It is clear that for larger values of $\gamma$, the
skew information reaches  its minimum values faster than that
shown for smaller values of $\gamma$. Also, for smaller values of
$\gamma$, $\mathcal{S}_I$ increases gradually but for larger
$\gamma$ the skew information increases hastily. This behavior is
clearly displayed  in Fig.(3a) for resonant case.  As one
increases the detuning by $30\%$, the behavior of $\mathcal{S}_I$
changes dramatically. The skew information increases and decreases
hastily regardless of the ratio $\gamma$. In this case the minimum
values of $\mathcal{S}_I$ is  reached much earlier i.e $T$ reduces
by$ 5\%$.

Fig.(5) describes the dynamics of $\mathcal{S}_I$  for different
values of $\gamma=2,4,6$ and it is assumed that the initial system
is prepared in resonant case. It is clear that, the behavior is
similar to that shown in Fig.(3b). This show that one can control
on the behavior of the skew information either by the detining
parameter or the ratio $\gamma$.

\section{Conclusion}
In this contribution we investigate one of the most quantity of
information, skew information, for a particle of CPB which is a
promising candidate for quantum information and computations. The
suggested model  consists  of a single Cooper pair Box prepared
initially in the excited state interacts with a cavity mode
prepared in the number state.

The effect of the cavity and the Cooper pair Box's parameters  on
the dynamics of the skew information  is investigated. Different
cases are considered the resonant and non-resonant cases. For
resonant case, the skew information increases gradually and takes
more time to reaches its maximum or minimum values. However  for
non resonant case, the skew information increases faster for
larger values of the detuning parameter and consequently the
number of oscillations between the maximum and minimum values
increases.

The number of photons inside the cavity has a noticeable effect,
where as one increases the number of photons, the skew information
increases faster and the oscillation increases. However, if we
increase the number of photon by $1\%$, the oscillations time
between the maximum and minimum values reduced by $2\%$.

The effect of the relative ratio of Josephson junction capacity
and the gate capacities plays in important role on the the
dynamics of skew information. For small values of this ratio the
oscillations of the skew information increases and consequently
the revival time decreases. On the other hand the skew information
increases faster for smaller values of the ratio of junction
capacity and the gate capacities and gradually for larger values
of this ratio.

{\it In conclusion:} it is possible to control on the dynamics of
the skew information by controlling on the cavity or the CPB's
parameters. Therefore one can speed up the skew information to
reach its maximum value either by increasing the number of photon
inside the cavity or considering non-resonant case with larger
detuning.

\end{document}